\documentclass[preprint]{iucr}              % DO NOT DELETE THIS LINE

     \journalcode{J}              % Indicate the journal to which submitted
\usepackage{graphicx}
\usepackage{amsmath}
\usepackage{amssymb,amstext}
\usepackage{todonotes}
\usepackage{ulem}
\renewcommand{\vec}[1]{\ensuremath{\boldsymbol{#1}}}
\newcommand{\mat}[1]{\ensuremath{\uuline{#1}}}

\usepackage{xcolor}
\usepackage{soul}

\begin{document}                  % DO NOT DELETE THIS LINE

     %-------------------------------------------------------------------------
     % The introductory (header) part of the paper
     %-------------------------------------------------------------------------

     % The title of the paper. Use \shorttitle to indicate an abbreviated title
     % for use in running heads (you will need to uncomment it).

\title{xrayutilities: A versatile tool for reciprocal space conversion of scattering data recorded with linear and area detectors}
\shorttitle{Q-conversion of 1D and 2D detector data}

     % Authors' names and addresses. Use \cauthor for the main (contact) author.
     % Use \author for all other authors. Use \aff for authors' affiliations.
     % Use lower-case letters in square brackets to link authors to their
     % affiliations; if there is only one affiliation address, remove the [a].

\cauthor[a]{Dominik}{Kriegner}{dominik.kriegner@gmail.com}{address if different from \aff}
\author[b]{Eugen}{Wintersberger}
\cauthor[a]{Julian}{Stangl}{julian.stangl@jku.at}{address if different from \aff}

\aff[a]{Institute of Semiconductor and Solid State Physics, Johannes Kepler University, Altenbergerstr. 69; A-4040 Linz, \country{Austria}}
\aff[b]{DESY, Notkestrasse 85, D-22607 Hamburg, \country{Germany}}

     % Use \shortauthor to indicate an abbreviated author list for use in
     % running heads (you will need to uncomment it).

\shortauthor{D.~Kriegner, et al.}

     % Use \vita if required to give biographical details (for authors of
     % invited review papers only). Uncomment it.

%\vita{Author's biography}

     % Keywords (required for Journal of Synchrotron Radiation only)
     % Use the \keyword macro for each word or phrase, e.g.
     % \keyword{X-ray diffraction}\keyword{muscle}

%\keyword{keyword}

     % PDB and NDB reference codes for structures referenced in the article and
     % deposited with the Protein Data Bank and Nucleic Acids Database (Acta
     % Crystallographica Section D). Repeat for each separate structure e.g
     % \PDBref[dethiobiotin synthetase]{1byi} \NDBref[d(G$_4$CGC$_4$)]{ad0002}

%\PDBref[optional name]{refcode}
%\NDBref[optional name]{refcode}

\maketitle                        % DO NOT DELETE THIS LINE

\begin{synopsis}
Algorithms for the reciprocal space conversion of linear and area detectors
implemented in an open-source Python package.
\end{synopsis}

\begin{abstract}
We present general algorithms to convert scattering data of linear and area
detectors recorded in various scattering geometries to reciprocal space
coordinates. The presented algorithms work for any goniometer configuration
including popular four-circle, six-circle and kappa goniometers. We
avoid the use of commonly employed approximations and therefore provide
algorithms which work also for large detectors at small sample detector
distances. A recipe for determining the necessary detector parameters including
mostly ignored misalignments is given. The algorithms are implemented in a
freely available open-source package.
\end{abstract}

     %-------------------------------------------------------------------------
     % The main body of the paper
     %-------------------------------------------------------------------------
     % Now enter the text of the document in multiple \section's, \subsection's
     % and \subsubsection's as required.

\section{Introduction}

Elastic X-ray scattering is a very widely applied technique to study the
structure of materials ranging from single crystals, powders and other forms of
hard condensed matter to biologic tissues and organic molecules.  Crystalline
as well as non-crystalline matter like liquids and amorphous materials can be
studied by techniques like X-ray diffraction and reflectometry. A variety of
approaches exist to analyze the scattering data. Some quantities can be
directly determined from the measurements (e.\,g. lattice
parameters~\cite{Bond1960,Fewster1999}, layer
thicknesses~\cite{Warren1969,Pietsch2004}). Other types of analysis involve
simulation of the scattering signal to determine strain and material
composition~\cite{Stangl2004,Wintersberger2010}, or the model-free
determination of real space structure using phase retrieval
algorithms~\cite{Fienup1982,Miao1998,Diaz2009,Minkevich2011}. All of those
approaches have in common that the analysis is most of the time performed in
reciprocal space, and hence require a conversion of experimentally measured
data into reciprocal space.  While the particular analysis steps differ for
different experiments, the conversion itself is a common step, which needs to
be performed for a lot of different techniques.  This has been treated very
well for the case of point detectors~\cite{Busing1967,Lohmeier1993,You1999,Bunk2004}.
For 1D and 2D detectors, which are used more and more frequently, several
issues related to the detector geometry and detector alignment complicate a
correct conversion. 

For some application like protein crystallography or powder diffraction,
experimental as well as analysis schemes are standardized. Examples are software
packages to extract peak positions and intensities in
protein-crystallography\cite{Leslie2006} and powder
diffraction~\cite{Lutterotti1999,Rodriguez2001,fit2d}, or the \emph{pdb}
protein data bank format and \emph{cif} crystallographic information file
format established by the International Union of Crystallography for exchange
of structure files and experimental data.

For most other cases, only very specialized solutions to particular experiments
exist, each containing solutions for some aspects important for the respective
case, e.\,g. the experimental control software \emph{spec}~\cite{certif} used
at various synchrotron sources is able to do reciprocal space conversion for
several goniometer geometries, however only for point detectors. Commercial
software supplied with several diffractometers is optimized for certain
geometry/detector combinations. 

A particular problem of 1D and 2D detectors are misalignments with respect to
the ideal case: At zero detector angle, the line or plane of the detector
should ideally be perpendicular to the incident X-ray beam. In practice, this
will not be the case, even if deviations will usually be very small. For
2D-detectors, in addition a rotation of the detector around the incident beam
direction can occur. In most cases, these misalignments will be unintentional
and rather small, they are difficult to measure and hence often neglected.
However, considering these effects is actually important to obtain correct and accurate
reciprocal space coordinates.

We present a general applicable algorithm for the conversion of experimental
data recorded with point, linear and area detectors, and for any diffractometer
with arbitrary number and sequence of axes. For this purpose, we generalize the
algorithms presented in several papers~\cite{Busing1967,Lohmeier1993,You1999,Bunk2004}
to arbitrary goniometer geometries {\it without approximations}: we consider the fact
that most 1D and 2D detectors are flat and hence the relation between channel
number and scattering angle is non-linear.  Furthermore, we provide recipes how to
determine the necessary detector parameters from a set of simple scans around
the primary beam. These scans also enable the user to determine the mentioned
and several other misalignment parameters. 

To keep our solution as general as possible, we have implemented it within a
freely available software package \emph{xrayutilities}~\cite{xrutils_web}.
This generalized algorithm is also particularly useful for automatized
tool-chains as planned by several synchrotron facilities right
now~\cite{passerelle}. In addition to the reciprocal space conversion described
in this article, \emph{xrayutilities} includes routines to read various data
formats and methods to calculate experimental parameters from material
properties. More information on those parts can be found on the
\emph{xrayutilities} web page~\cite{xrutils_web}. This article focuses on the
reciprocal space conversion part: After the introduction of the applied
algorithms following the approach of You, 1999 we show the extension for linear
and area detectors and explain the use of our algorithms and how necessary
parameters can be determined. We demonstrate the application of our approach
for few selected examples of both laboratory diffractometers and synchrotron
beamlines. In the appendix we discuss one complete example, including the
particular entries into the \emph{xrayutilities} package to define the
diffractometer geometry, correctly initialize a 2D detector setup, and
convert a 2D detector frame into reciprocal space.  

\section{Angular to reciprocal space conversion}

Conversion of angular coordinates to reciprocal space can be tedious since one
needs special equations for every diffractometer/detector geometry. For several
diffraction geometries explicit formulas are given for example in Pietsch,~2004.
However the conversion can be performed in a general way as long as the
information about the goniometer geometry is available together with the
experimental angles. The algorithm presented below therefore needs not just the
experimental angles as input parameters but also the description of the
goniometer.  To work for arbitrary goniometers the physical order of the
cradles, i.\,e. how they are mounted on each other, as well as the orientation
of the rotation axes are needed. An example is given below. To unambiguously
specify the rotation axes of the goniometer circles, a reference coordinate
system is fixed to the laboratory frame. It is useful to choose this coordinate
system in a way that the primary X-ray beam propagates along one of the
coordinate axes.

The reciprocal space coordinates we want to know for each measured point are
coordinates of the momentum transfer $\vec q = \vec k_f - \vec k_i$, where
$\vec k_i$ is the wave-vector of the incident X-ray beam, $\vec k_f$ the
wave-vector of the scattered beam towards a particular detector (pixel)
position. The coordinates of $\vec q$ we are interested in are those in a
coordinate system fixed to the sample under investigation. The coordinates of
$\vec k_{i}$ in the laboratory system are fixed by our choice of the coordinate
system. The coordinates of $\vec k_f$ are given by the angle positions of the
detector arm. To describe $\vec q$ in the sample coordinate system, we also
need all goniometer angles changing the sample orientation.

\begin{figure}
 \caption{Two dimensional sketch of the scattering process and the coordinate
 systems attached to the laboratory ($x_L$,$y_L$) and sample ($x_s$,$y_s$).
 Shown are the incidence and exit wave vectors $\vec k_{i,f}$ as well as the
 momentum transfer of the scattering process $\vec q$. The laboratory
 coordinate system is chosen to have the $\vec{\hat x}$ direction to coincide
 with the primary beam direction.}
 \label{fig:scattering_process}
 \includegraphics[scale=1.3]{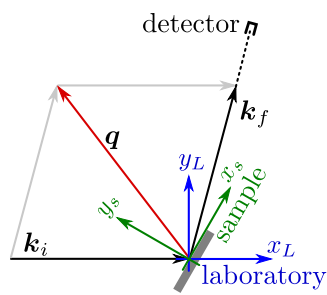}
\end{figure}

A minimal two dimensional example illustrating the definition of our
laboratory coordinate system (blue) is shown in
Fig.~\ref{fig:scattering_process}.  The coordinate system attached to the
sample is indicated in green in Fig.~\ref{fig:scattering_process}.  So what we
have to deal with are coordinate transformations between the different involved
coordinate systems. Those coordinate transformations can be written as matrix
equations. We will reproduce some of the essential equations of earlier
papers~\cite{Busing1967,You1999} so that reader can follow the further
generalization for one and two dimensional detectors. We consider just a point
detector for the moment, the generalization for one or two dimensional
detectors is shown in Sec.~\ref{sec:more_dimensions} below.

The sample coordinate system we have been talking about is actually the
coordinate system attached to the innermost goniometer circle (because this is
what the goniometer angles describe). Of course, the sample can in addition
have a certain orientation with respect to this circle. This is most evident in
the case of a crystalline sample, where the directions of the reciprocal space
of the crystal have a certain orientation with respect to the sample holder.
This and the coordinates of the momentum transfer in the reciprocal space of
the crystal are described below by matrices $\mat U$ and $\mat B$,
respectively.  In the following, we describe this ''most complicated'' case of
a single crystalline sample. For amorphous or powder samples etc., $\mat U$
and/or $\mat B$ can be replaced by identity matrices.

In the crystal lattice a momentum transfer might be described by the column vector
\begin{equation}
\vec h_c = \sum_{j=1}^{3} h_j \vec b_j = \mat B \vec h
\label{eq:hc}
\end{equation}
with $\vec b_j$ the reciprocal lattice vectors or the matrix $\mat B$ formed
from those vectors and $h_j$ generally referred to as $h$, $k$, $l$. In case
$h$, $k$, $l$ are integer values they are called Miller indices. An explicit
representation of $\mat B$ is for example given in~\cite{Busing1967}. 

To connect the momentum transfer in the crystal with the momentum transfer in
the coordinate system attached to the sample goniometer the orientation matrix
$\mat U$ is introduced.
\begin{equation}
\vec h_s = \mat U \vec h_c = \mat U \, \mat B \vec h
\label{eq:hs}
\end{equation}
This momentum transfer converted to the laboratory frame is equal to $\vec q$.
The conversion of the momentum transfer is achieved by another coordinate
transformation (described by matrix $\mat S$), which solely depends on the
sample orientation and therefore the goniometer angles which move the sample.
\begin{equation}
\vec q = \vec k_f - \vec k_i = \mat S \vec h_s
\label{eq:conversion_condition}
\end{equation}
In the case of integer $h$, $k$, $l$, the former
equation basically is the Laue equation.

Also any detector rotation can be expressed as coordinate transformation
described by $\mat D$ and therefore the exit wave vector is given by
\begin{equation}
\vec k_f = \mat D \vec k_i
\label{eq:kf}
\end{equation}

Combining Eqs.~\ref{eq:hs} to~\ref{eq:kf} one obtains the momentum transfer in
the crystal coordinate system
\begin{align}
 \vec h_c = \left( \mat S \, \mat U \right)^{-1} \left(\mat D - \mat 1 \right) \vec k_i
 \label{eq:qconversion}
\end{align}
with the identity matrix $\mat 1$.  Note that all of the used coordinate
transformations are transformations between two orthogonal coordinate systems,
and therefore yield orthogonal matrices, which are invertible.  The matrix
$\mat B$, which is formed from the reciprocal space vectors $\vec b_j$, which
are linear independent is also invertible. Since the above conversion involves
a matrix inversion this is important for the algorithm to work.

The rotation matrices $\mat S$ and $\mat D$ can be deduced from the description
of the goniometer by multiplying the rotation matrices from each circle
starting with the outer most circle.

\begin{align}
 \mat D, \mat S = \text{(outer most)} \cdot \dots \cdot \text{(inner most)}
 \label{eq:matrix_order}
\end{align}

For that purpose the goniometer rotation axis needs to be defined in the
laboratory coordinate system for the case when all circles are set to zero. It
is therefore useful to choose the coordinate system in a way, which allows this
description to be as simple as possible. Keep in mind that later also the
detector directions need to be determined in this coordinate system. In
addition to the direction of the rotation axis also the rotation sense needs to
be described. We use the mathematical definition of rotation sense.  For most
goniometers (except for special geometries like $\kappa$-goniometer, treated
separately below) the rotation axis point along primitive directions. When
looking on the rotation from the positive side of the corresponding direction,
clockwise rotation is left-handed, i.\,e., negative, and an anticlockwise
rotation is right-handed (positive). For example, we will call a
clockwise/left-handed/negative rotation of angle $\alpha$ around the $x$-axis a
'{\tt x-}'-rotation, described by the following rotation matrix:

\begin{align}
 \mat R_{x-}(\alpha) = \begin{pmatrix} 1 & 0 & 0 \\ 0 & \cos \alpha & \sin \alpha \\ 0 & -\sin\alpha & \cos \alpha \end{pmatrix}
 \label{eq:rotmat}
\end{align}

\subsection{example of a goniometer definition}

\begin{figure}
 \caption{sketch of a goniometer with three sample axis, and two detector
 rotations. A red and a blue plane indicate the detector rotation planes of
 the inner detector rotation for two different positions of the outer detector
 arm rotation.}
 \label{fig:goniometer}
 \includegraphics[scale=0.12]{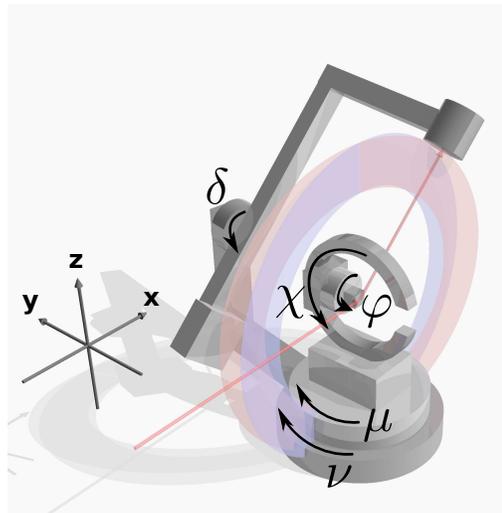}
\end{figure}

To elucidate the definition of a goniometer we use the goniometer shown in
Fig.~\ref{fig:goniometer}. The goniometer has a 3S+2D geometry, which means it
offers three degrees of rotation for the sample and two independent degrees of
freedom for the detector. The goniometer axes are specified by their axis
direction and rotation sense. The coordinate system is chosen to have $x$
pointing downstream along the primary beam, $z$ is pointing upwards and $y$
pointing backwards to have a right-handed reference frame.  In
\emph{xrayutilities} we describe each rotation axis with one character giving
the axis direction (either {\tt x}, {\tt y}, or {\tt z}) and the rotation sense
(either {\tt +} or {\tt -}). This description needs to be supplied for the
sample and detector circles for the case where all axis are at zero position
starting with the outermost circle. The goniometer in Fig.~\ref{fig:goniometer}
has three sample circles ($\mu$, $\chi$, $\varphi$) with the indicated rotation
directions. The outermost angle $\mu$ turns clockwise around the $z$-coordinate
axis and thus is described by '{\tt z-}'.  The complete sample goniometer is
described by the following set of rotations: ('{\tt z-}','{\tt x-}','{\tt
y+}'). For the detector circles turning around the $z$ and $y$-directions, the
corresponding definition is ('{\tt z-}','{\tt y-}'). A full example of how to
insert these definitions into {\it xrayutilities} is given in the appendix.

\subsection{special rotation directions (Kappa goniometer)}
\label{sec:kappa_goniometer}

In addition to rotations around axis of the coordinate system an often used
goniometer geometry is the so called
$\kappa$-geometry~\cite{Poot1972,Thorkildsen1999}, in which one of the rotation
axis has an angle of typically 45 to 60$^\circ$ with one of the other axis.
In \emph{xrayutilities} we define such an axis by {\tt k+} or {\tt k-}. The
plane of the $\kappa$-rotation axis as well as its angle with respect to a
reference direction are specified in a configuration file by the options {\tt
kappa\_plane} and {\tt kappa\_angle}. The rotation matrix needed in matrix
$\mat S$ can be determined easily using an general equation given in the
appendix.

\section{Angular to momentum space conversion for 1D and 2D detectors}
\label{sec:more_dimensions}

Equation \ref{eq:qconversion} describes the conversion from angular coordinates
of a general goniometer to reciprocal space for a point detector only. The
generalization for a linear or area detector requires information about the
pixel size and distance, and the direction into which a pixel row extends.
Usually in the data files only one angular position is stored for every data
point recorded with a multidimensional detector. This angular coordinate
corresponds to the position of the so called center channel/pixel ($n_0$),
which is the pixel hit by the primary beam when all the angles are set to zero.
For all other detector pixels their position needs to be determined.  This is
easiest for the case of a curved one dimensional detector in which every
detector channel or pixel covers the same detector angle segment. Every
detector pixel (identified by the channel number $n$) corresponds to a detector
angle $2\theta$ given by
\begin{equation}
2\theta(n) = 2\theta_0 + (n-n_0)/N
\label{eq:curved_1ddet}
\end{equation}
where $N$ is the number of channels per unit of angle of the detector circle.
When the detector angle is expressed in degree $N$ equals the number of
channels per degree of rotation.

\begin{figure}
 \caption{Sketch of a linear detector mounted at distance $L$ from the
 center of rotation of the goniometer. In (a) the detector direction specifying
 the direction along which higher channel numbers are found is given by
 $\vec{\hat d}$. Also indicated is the width of one channel $w$ and the center
 channel $n_0$, which is the channel where the primary beam is centered at zero
 detector angle. Panel (b) shows the possible detector tilt $\beta$
 (misalignment), resulting if a detector is not mounted perfectly perpendicular
 to the X-ray beam.}
 \label{fig:detector_chpdeg}
 \includegraphics{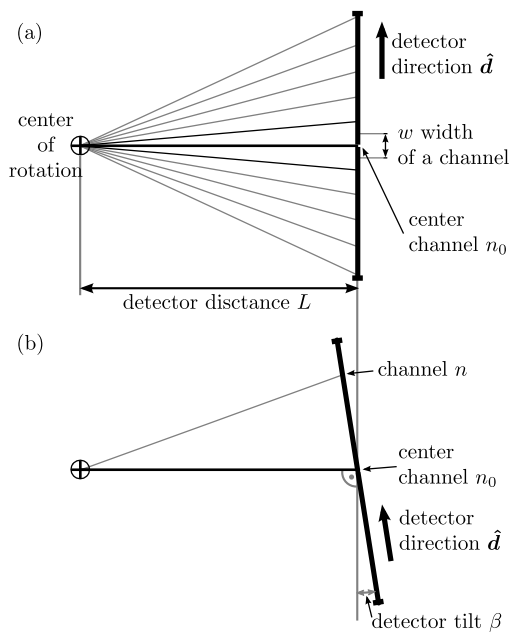}
\end{figure}

Most modern detectors are, however, straight as shown in
Fig.~\ref{fig:detector_chpdeg} and \emph{not} curved and therefore
Eq.~\ref{eq:curved_1ddet} is not generally applicable. Only in the limit of a
large sample detector distance the curved and straight detectors become
indistinguishable. Nevertheless the channel per degree approximation is
frequently used in practise.  In \emph{xrayutilities} one-dimensional detectors
are treated as straight detectors and Eq.~\ref{eq:qconversion} is adjusted
accordingly. For each detector pixel $n$, the corresponding direction of a
scattered beam hitting this pixel is calculated, replacing Eq.~\ref{eq:kf} by
\begin{equation}
\vec k_f(n) = \left| \vec k_i \right| \mat D \left(\vec{\hat k}_i + \vec{\hat d} (n-n_0) w/L\right)/\| \vec{\hat k}_i + \vec{\hat d} (n-n_0) w/L\|,
\label{eq:kf_1d}
\end{equation}
where $w$ and $L$ are the size of a detector pixel and distance from sample to
detector as shown in Fig.~\ref{fig:detector_chpdeg}. The direction in which the
detector channel number increases is given by $\vec{\hat d}$. A 'hat' on a
vector indicates a unit vector. The fraction $w/L$ is in the case of large
sample to detector distance equal to \( 1/N \), where $N$ is is the number of
channels per radians, and Eq.~\ref{eq:kf_1d} effectively simplifies to the form
of Eq.~\ref{eq:curved_1ddet}. For a two dimensional detector with channel
directions $\vec{\hat d}_1$ and $\vec{\hat d}_2$ we can write an equivalent
equation for the exiting wave vector of channel ($n_1$,$n_2$) including the
width of the detector pixels $w_{1,2}$ as

\begin{equation}
\vec k_f(n_1,n_2) = \left| \vec k_i \right| \mat D \left(\vec{\hat k}_i + \vec{\hat d}_1 (n_1-n_1^{(0)}) w_1/L + \vec{\hat d}_2 (n_2-n_2^{(0)}) w_2/L\right)/\| \cdots \|,
\label{eq:kf_2d}
\end{equation}

Using the description of the detector in real space we do therefore avoid the
''channel per degree'' approximation, which implicitly assumes a detector is
curved and therefore does not work for small sample detector distances with
large detectors. Inserting Eqs.~\ref{eq:kf_1d} and~\ref{eq:kf_2d} into
Eq.~\ref{eq:hs} and~\ref{eq:conversion_condition} yields general equations for
the reciprocal space conversion of linear and area detectors.  The possible
misalignments are included in those equations via the pixel directions
$\vec{\hat d}$ for the 1D and $\vec{\hat d}_{1,2}$ for the 2D detectors.

\subsection{Detector parameters of 1D detectors}
\label{sec:det_param1d}

\begin{figure}
 \caption{Determination of the ratio $w/L$ and center channel for a linear
 detector with pixel size of 50~$\mu$m from the variation of the beam position
 during a scan through the primary beam by fitting Eq.~\ref{eq:nlintilt} to the
 recorded data. Shown are the channel numbers at which the beam is observed vs.
 the detector angle during the scan. The lower plot shows the experimental data
 as well as the fits when the linear trend is subtracted.  A fit with (full
 line) and without (dashed line) considering a detector tilt is shown for two
 detector distances of 250~mm and 380~mm.}
 \label{fig:detector_param}
 \includegraphics[scale=0.5]{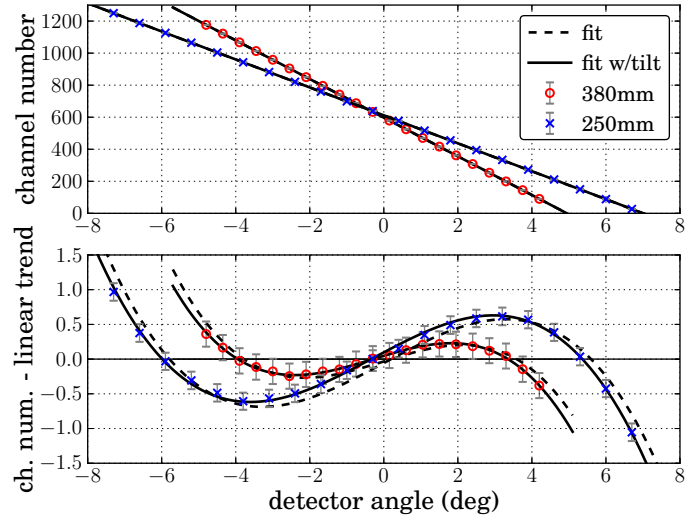}
\end{figure}

For the conversion algorithms described above several detector parameters are
needed. Among them the detector distance $L$ and width of one detector channel
$w$. Since neither the detector distance nor the width of one channel can be easily
measured very accurately, we use the fact that only their ratio is
needed, which can be determined from an angular scan through the primary beam.
Assume a linear detector is mounted along the direction in which the detector
arm of the used instrument moves. Scanning the detector angle will move the
primary beam over the detector, by modeling this movement we are able to
determine the needed quantities. Therefore assume a linear detector mounted at a distance
$L$ like the one shown in Fig.~\ref{fig:detector_chpdeg}. Neglecting for the
moment a possible detector tilt, i.\,e. when a detector is not mounted
perfectly perpendicular to the X-ray beam, the detector channel at which the
detector is hit for a detector arm angle $2\theta$ is given by
\begin{equation}
n(2\theta) = \frac{L}{w} \tan 2\theta + n_0.
\label{eq:nlin}
\end{equation}
If a detector tilt $\beta$ as shown in Fig.~\ref{fig:detector_chpdeg} is
included above equation needs to be modified and one finds
\begin{equation}
n(2\theta,\beta) = \frac{L}{w} \frac{\sin 2\theta}{\cos\left(2\theta -\beta\right)} + n_0.
\label{eq:nlintilt}
\end{equation}
By fitting one of the two models one can find the necessary detector parameters
needed for the reciprocal space conversion of a linear detector. For this
purpose a scan through the primary beam should be performed with the linear
detector and the detector spectra should be saved at every position. Since not
only the slope is determined (which only needs two spectra), it is required to
acquire several spectra, we suggest typically $\gtrsim 10$.  For the
determination of the parameters two functions are provided in
\emph{xrayutilities}. One of them automatically processes the spectra of a
linear detector acquired during a scan through the primary beam and determines
the position of the primary beam in every spectrum by fitting a Gaussian peak.
From this fitting the position of the primary beam is found with sub-pixel
precision. The second function needs the user to determine the channel number
of the primary beam and supply it to the function, which should be used only in
cases the first function is not applicable. Calling one of the two functions
will produce a plot as the one shown in Fig.~\ref{fig:detector_param}, which
shows the channel number at which the beam was hitting the detector together
with the variation expected from the model(s). A second subplot shows the
comparison of model and measured data with the linear trend subtracted, for two
different detector distances of 380~mm and 250~mm for a straight linear
detector with a pixel size of 50~$\mu$m. The different distance shows up as
different slope in the upper subplot. When the linear trend is subtracted the
non-linearity due to the trigonometric functions in Eqs.~\ref{eq:nlin}
and~\ref{eq:nlintilt} can be seen.  The fit is also sensitive to a tilt of the
detector from the direction perpendicular to the primary beam as can be seen
from the comparison of the model without tilt (Eq.~\ref{eq:nlin}, dashed line)
and model with tilt (Eq.~\ref{eq:nlintilt}, full line). For the measurements
shown in Fig.~\ref{fig:detector_param} an artificial tilt of 0.3~deg was
introduced, which was also determined by the fit. Note that a tilt of the
detector is not only a result of a not perfectly mounted detector, but can also
result from the fact that experimentalists choose to use a center channel,
which does not correspond to the true center of the detector and do this by
redefining the zero position of the detector arm rotation, which introduces a
tilting of the detector by exactly the angle by which the detector arm angle
changed!  Using a translation of the detector perpendicular to the beam
(mounted on top of the detector arm) would prevent such a tilt, is, however,
very often not available.

\begin{figure}
 \caption{Six reciprocal space maps recorded around the Si (331) Bragg peak
 using the same sample movement, however, using different parts of a linear
 detector for detection. The white cross marks the nominal position of the Si
 (331) Bragg peak.  Panels (a,d) were recorded when the detector was offset to
 lower angles, panels (c,f) with an offset to higher angles, whereas panels
 (b,e) were recorded with the signal centered on the detector. For panels (a)
 to (c) the exact reciprocal space conversion described in the text was used,
 whereas for panels (d) to (f) the ''channel per degree'' approximation was
 applied leading to errors in the observed peak positions.}
\label{fig:si331maps}
\includegraphics[scale=0.15]{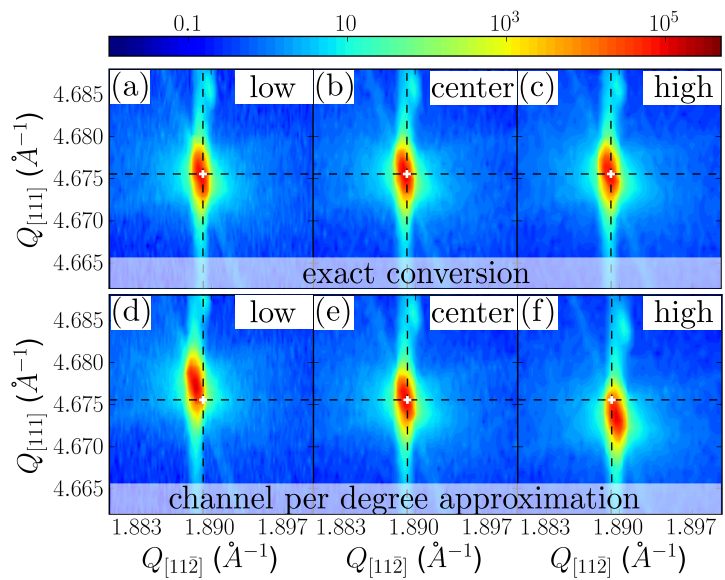}
\end{figure}

Basically this shows the necessity of non-linear models instead of the simpler
linear fitting, which is a reasonable approximation only in the case of a large
sample to detector distance. If a linear or area detector covers an angular
range of $\gtrsim 8^\circ$ it gets absolutely necessary to use the above
treatment since deviations would exceed one channel for typical channel sizes
of around 50 $\mu$m. Using large linear detectors at moderate sample detector
distances this limit is frequently reached especially in modern laboratory
diffractometers. To further highlight the necessity of the non-linear models we
show an example of a x-ray diffraction reciprocal space map measurement of the
Si (331) Bragg peak of a Si(111) oriented single crystal in
Fig.~\ref{fig:si331maps}. The measurement was performed with the above
mentioned linear detector at a distance of 250~mm at a laboratory
diffractometer with CuK$\alpha$1 radiation produced by a Ge(220) channel cut
monochromator. The same measurement was repeated three times, however,
different parts of the detector were used for the detection of the diffracted
signal. We compare the reciprocal space maps obtained using our exact
conversion formalism in comparison with the ''channel per degree''
approximation which assumes a curved detector like described by
Eq.~\ref{eq:curved_1ddet}. Using this approximation (panels (d) to (f) in
Fig.~\ref{fig:si331maps}) we find that only the measurement using the central
part of the detector gives the correct peak position in reciprocal space/lattice parameter.  The
measurements performed with the detector offset to higher or lower angles would
result in a lattice parameter wrong by approximately 0.04\%, which is far
beyond the usual sensitivity of the experimental setup, which is said to be
$<1\times10^{-4}$ and can be increased further as for example outlined
in~\cite{Fewster1999}. Using the accurate conversion no such shifts are
observed and the three measurements (panels (a) to (c)) are indeed
equivalent.

\subsection{Detector parameters of 2D detectors}
\label{sec:det_param2d}

For 2D detectors similar determination of the $w_{1,2}/L$ ratio can be
performed if the detector rotation and other misalignments (see below) are
negligible, by decomposing the problem into two separate one dimensional
problems. In case the detector is rotated around the primary beam the problem
can no longer be decomposed. In particular one more problem specific to two
dimensional detectors arises. Very often the true zero of the outer detector
arm rotation is not known. For the inner detector rotation this does not imply
any particular problem since it shows up only as additional tilt, which can be
easily corrected.  However, an offset in the outer detector rotation implies a
rotation of the {\it rotation axis} of the inner detector rotation. In case the
outer motor is not at the true zero the inner rotation will no longer rotate
perpendicular to the primary beam as indicated by the two rotation planes shown
in Fig.~\ref{fig:goniometer}. In case the outer motor would be at its true zero
the detector would rotate in the blue plane, due to an offset in the outer
rotation the detector instead rotates in the red plane.  This means the offset
of the outer motor needs to be determined from alignment scans as well.

\begin{figure}
 \caption{Determination of detector parameters. Shown is the optimization error
 vs. the eight considered parameters: the center channels of the detector and
 the width of the pixels in both directions, the detector tilt (azimuth and
 tilt angle), the detector rotation and the offset of the outer goniometer
 stage. The found optimum is marked by a black circle. All other points correspond
 to fits which did not reach the global minimum, those were produced from
 various different starting parameters of the fit. Green lines are guides to
 the eye to visualize the minimum found in all the parameters. The points of
 the fits are colored to enable the identification of correlations.}
\label{fig:detparam}
\includegraphics[scale=0.1]{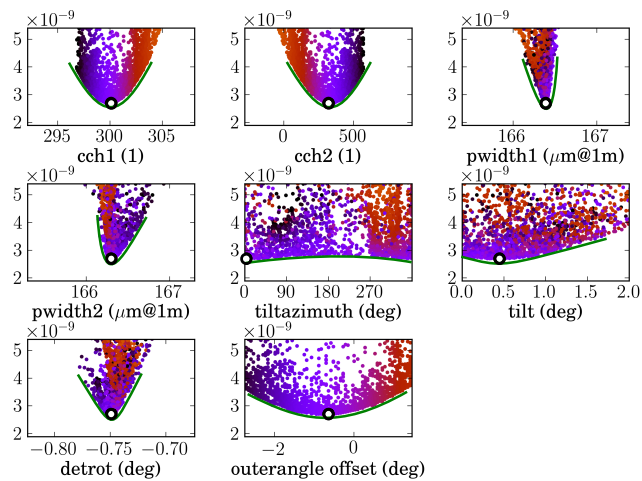}
\end{figure}

In general one needs to determine 8 parameters for a 2D detector: the center
channels ($n_1^{(0)}$, $n_2^{(0)}$), the ratios $w_{1}/L$ and $w_{2}/L$, and
directions of the vectors $\vec{\hat d}_{1,2}$ which are specified by the tilt,
tilt direction and rotation of the detector around the primary beam and the
outer angle offset described above. The untilted directions of the vectors
$\vec{\hat d}_{1,2}$ can be determined by knowledge of the primary beam and
detector rotation directions unambiguously and do therefore not need to be
included in the fit. Similar as for the one dimensional detector those
parameters can be determined from scans through the primary beam. It is
necessary to use at least two scans, one with the inner and one with outer
detector arm rotation with sufficient points around the primary beam and
acquire a detector image at several positions. From the primary beam position
observed in those images the detector directions and other necessary parameters
can be determined by a fitting routine. As quality criterion of the fit the
average $\|\vec q\|$-value of the pixel at which the primary beam is observed
is used. This position is calculated using Eqs.~\ref{eq:conversion_condition}
and~\ref{eq:kf_2d}.  Since it is the primary beam we observe in all the images,
this $\|\vec q\|$-value should be zero when the correct detector parameters are
found.  \emph{xrayutilities} provides a function which takes the detector
images and performs a fit of the 8 described parameters. Since several of those
parameters, e.\,g. the offset of the outer motor with one of the center
channels, are coupled with each other, the fit is performed in a way that it
starts from several starting parameters to find the global minimum in the
parameter space. An example of such a fit is shown in
Fig.~\ref{fig:detparam}.
The detector parameters have been determined from two perpendicular scans
through the primary beam using a Maxipix-detector with effectively 516 times
516 pixels at beamline ID01/ESRF~\cite{esrfid01}. One scan was performed using
the outer detector arm motor (moving horizontal) and the other one with the
inner detector circle (moving vertical). In each scan we used 70 points, which
means that in total 140 images were used. Note that using considerable less
number of points does reduce the quality of the fit until no reliable
statement can anymore be made about the misalignment parameters. It is
therefore suggested to use a comparable number of points as in this example.
Furthermore the images were manually selected to use just those with the full
primary beam in active area of the detector.  These images were fed into the
algorithm described above and the detector rotation, tilt, tilt azimuth and
outer angle offset along with the center channels and detector pixel size were
determined. This determination is shown in Fig.~\ref{fig:detparam}, were the
average $\|\vec q\|$-deviation of the detector pixels positions associated with
the primary beam in the performed scans is shown. The value of this deviation
is approximately the offset of the absolute value of the momentum transfer in
reciprocal space. This deviation shows a clear minimum in all the 8 parameters.
These 8 parameters are: 
1/2. {\bf cch1,2} are the center channels of the detector in both directions.
This is the pixel number where the primary beam is hitting the detector at the
true zero position of the detector arm (including the outer angle offset).  
3/4. {\bf pwidth1,2} are the pixel widths of the pixels in the two detector
directions. The unit of these values in the plot is the size of the pixels in
$\mu$m assuming a detector distance of 1~m.  This corresponds to the parameter
$w/L$ from above. If the pixel size is known the detector distance can be
calculated or vice versa.  
5. {\bf tiltazimuth} is an angle giving the direction of the detector tilt. An
value of 90 and 270~degrees corresponds to a tilt rotation around the first
pixel direction and 0 and 180 around the second pixel direction, respectively.

6. {\bf tilt} is the tilt angle of the detector plane around the direction
given by the tiltazimuth, since the tilt-azimuth runs from 0 to 360 only
positive tilts are used.  
7.  {\bf detrot} is the detector rotation around the primary beam direction in
degrees.  
8. {\bf outerangle offset} is the offset of the outer detector arm rotation in
degrees. 

For the determination of those parameters fits with various different starting
parameters are performed. This is absolutely mandatory since several of the
parameters are correlated, and therefore a single fit would not necessarily
find the global optimum. One correlation, which can be easily imagined is the
correlation of one of the center channels with the outer rotation offset. Since
the detector orientation is not given to the script and automatically
determined from the given scan data it is not apriori clear if this is center
channel 1 or 2.  In Fig.~\ref{fig:detparam} this correlation can be seen in the
plot showing center channel 2 (cch2) and the outer angle offset. The cloud of
points from center channel 2 is rather broad since this value changes when the
outer angle offset is changed. In fact the coloring of the cloud of points of
center channel 2 reveals that it is only mirrored with respect to the one of
the outer angle offset. A not so intuitive correlation is the one of the
detector tilt with the outer angle offset. Imagine a not tilted detector,
which is however offset in the outer angle. Such an offset will effectively
tilt the detector by exactly the offset in the outer angle with a tilt azimuth
of 90 or 270~deg. In case the detector is already slightly tilted in an
arbitrary direction without outer angle offset, the tilt due to an outer angle
offset will overlay with the initial tilt and influence the tilt azimuth and
tilt in a non-trivial manner. The optimal set of parameters for the area
detector of beamline ID01 was determined from the point with the lowest lowest
error (global minimum in the parameter space) of~$2.70\times10^{-9}$. 

To elucidate the benefit of considering those ``misalignment'' parameter, which
are usually omitted, we also give the obtained errors when several of the quantities
are fixed. When only the center channels and pixel widths are fitted we obtain
an error of $1.68\times10^{-6}$, which is orders of magnitude higher than our
optimal error. In the present example the most important parameter is the detector
rotation which brings the error down to $3.65\times10^{-9}$. The second most
important parameter is the outer angle offset, which brings down the error
further to $2.81\times10^{-9}$. If only the tilt (tilt azimuth and tilt angle)
or the outer angle offset are fitted without the detector rotation, the error
can only be reduced by less than 10\% from the value without considering any
misalignment. Only when all parameters are considered in the fit the error
can be reduced to the optimum. This clearly shows that all those parameters
should be considered to obtain the correct reciprocal space positions of the
full area detector.

It should be noted that some of the parameters like the offset of the outer
detector arm rotation and detector tilt can only be determined when the
detector is spanning a certain angular range, the user is therefore urged to
check the resulting plots in order to see if the parameters are reasonable.
In cases where the detector distance is large the outer angle offset can only
be determined with large error and it might be better to fix the corresponding
parameter during fitting.  Also the in cases when the detector tilt is small
the tilt azimuth will not have a clear minimum and is therefore a
indeterminable parameter.  The script also outputs the code needed for
initialization of the area detector, which considers the determined tilts and
rotations as shown in the appendix~\ref{sec:examplescript}. These detector
tilts of linear and area detectors are then included in the reciprocal space
conversion. When the detector pixel position is calculated as described in
Eq.~\ref{eq:kf_1d} and~\ref{eq:kf_2d} the detector direction $\vec{\hat d}$ or
$\vec{\hat d}_{1,2}$ need to be rotated accordingly before the detector
position is calculated. 

In case no detector rotations are available, e.\,g.  when using an area
detector for powder diffraction we refer to the fit2d software~\cite{fit2d} for
determining the necessary parameters.

\section{Conclusions}

In conclusion we present algorithms for reciprocal space conversion for X-ray
diffraction experiments. We generalize the equations given
by~\cite{Busing1967,You1999} for the use of linear and area detectors. Using
our approach we can convert angles from arbitrary goniometers to reciprocal
space coordinates. For the conversion of linear and area detectors several
detector parameters, including all possible misalignments are needed. Recipes
were presented how those parameters can be determined from alignment scans for
both linear and area detectors. The software package including the presented
algorithms is available at {\tt http://xrayutilities.sourceforge.net}. The
algorithms were shown to determine the detector parameters of a 1D and 2D
detectors, which are otherwise not determinable.

%     % Appendices appear after the main body of the text. They are prefixed by
%     % a single \appendix declaration, and are then structured just like the
%     % body text.

\appendix

\section{Implementation details}

The software package is available from {\tt
http://xrayutilities.sourceforge.net} and is mainly coded in the popular
scripting language Python\texttrademark\ with some performance critical parts
written internally in the C programming language. The user needs to use only
the Python package. Installation is possible on all major platforms (Windows,
Mac, Linux/Unix). In addition to the reciprocal space conversion the package
offers several functions to read data from {\it spec}-files, {\it
spectra}-files and from laboratory diffractometers of Panalytical ({\it xrdml})
and Seifert ({\it nja}) as well as several CCD-data files ({\it edf},{\it
tiff}) and other function, which are described in the documentation found at
the web-page.

The reciprocal space conversions performance critical part, which implements
Eq.~\ref{eq:qconversion} and equivalents for linear and area detectors is coded
in C. As can be seen from Eq.~\ref{eq:qconversion} it mainly consists of
$3\times3$ matrix operations. For the conversion of angular positions the
matrices of the sample rotations need to be set up, for which the rotation
matrices as the one shown in Eq.~\ref{eq:rotmat} are used. These matrices are
multiplied in the order as given in Eq.~\ref{eq:matrix_order}. Furthermore,
the conversion involves one matrix inversion, which is performed using the
adjugate matrix formula. For matrix $\mat A$ this means the inverse is given by
\begin{align} 
\mat A^{-1} = \frac{\operatorname{adj} \mat A}
{\operatorname{det} \mat A} \text{ .} 
\end{align}

Depending on the number of sample ($n_s$) and detector circles ($n_d$)
to be considered in the conversion for a point detector $n_s+n_d+2$ matrix
multiplications, the matrix inversion, and one matrix-vector multiplication as
well as one matrix-matrix subtraction need to be performed for every data
point.

However, for linear and area detectors not for every detector pixel all the
conversion needs to be done. For one spectrum or image all the sample and
detector angles are considered to be constant and therefore the $n_s+n_d+1$
matrix multiplications as well as the matrix inversion need to be performed
only once. Only two matrix-vector multiplications and some vector arithmetic to
set up the pixel position need to be performed for every detector channel/pixel
which enables a fast conversion also for large area detectors. Furthermore,
the conversion for multiple data points is independent and can therefore be
easily performed in parallel on modern multiprocessor computers.

\section{example script for the reciprocal space conversion}
\label{sec:examplescript}

In the following an example script for the conversion of an image from an area
detector recorded using the goniometer shown in Fig.~\ref{fig:goniometer} is
given. This goniometer has the sample angles $\mu$, $\chi$ and $\phi$ and
detector angles $\nu$ and $\delta$. The rotation axis of this goniometer are as
given in the main text (sample circles: {\tt ('z-','x-','y+')} and detector
circles: {\tt ('z-','y+')}) and the primary beam propagates along positive
$x$-direction. The script is written in the Python\texttrademark\ script
language.

\begin{verbatim}
import xrayutilities as xu

Nav = (1,1) # number of pixel to average over (reduce amount of data)
region_of_interest = (100,500,100,500) # region of interest on detector
en = 9000 # xray energy in eV

qconv = xu.QConversion(('z-','x-','y+'),('z-','y-'),(1,0,0))
# 3S+2D goniometer: sample rotations: mu,chi,phi: ('z-','x-','y+');
# detector rotations: nu,delta: ('z-','y-')
# primary beam direction (1,0,0)

Si = xu.materials.Si
hxrd = xu.HXRD(Si.Q(1,0,0),Si.Q(0,1,0),en=en,qconv=qconv)
hxrd.Ang2Q.init_area('z-','y+',cch1=300.11,cch2=320.78,Nch1=516,Nch2=516, 
    pwidth1=1.6639e-04,pwidth2=1.6630e-04,distance=1.,detrot=-0.749,
    tiltazimuth=3.0,tilt=0.448,Nav=Nav,roi=region_of_interest)
outerangle_offset = -0.643

# experimental goniometer angles
mu,chi,phi = (20,0,0)
nu,delta = (40,0)

# conversion to Q
qx,qy,qz = hxrd.Ang2Q.area(mu,chi,phi,nu,delta,delta=(0,0,0,outerangle_offset,0))

# conversion to (h,k,l)
h,k,l = hxrd.Ang2HKL(mu,chi,phi,nu,delta,delta=(0,0,0,outerangle_offset,0),
                     mat=Si,dettype='area')
\end{verbatim}

All the detector parameters (like center channels, number of pixels, size of
the pixels and the distance as well as possible tilts and rotations of the
detector) need to be given to the $\tt init\_area$-function. This parameters
are the result of a fit like the one shown in Fig.~\ref{fig:detparam}.  The
conversion is then performed by calling the {\tt area}-method of an experiment
class object, where also the outerangle offset needs do be considered.  An
experiment object holds information about the experiment like orientation of
the crystal on top of innermost sample circle, the goniometer geometry and the
X-ray energy.  Here the object is initialized for an experiment performed using
X-rays with an energy of 9000~eV and silicon crystal with the [100] direction
along the primary beam at zero position and [010] along the $z$-axis.  The
second argument [010] gives the reference direction perpendicular to the
primary beam in the plane spanned by the innermost detector direction.  Dummy
values for the reference directions can be used when the {\tt Ang2HKL} function
is not used!

After execution of this script the variables {\tt qx, qy} and {\tt qz} hold the
reciprocal space position of the detector pixels specified by the region of
interest variable. In the case above those variables will have a shape of
400$\times$400, holding the $q$-position of the pixels 100 to 500 in both
detector directions. The variable {\tt Nav} could be used to reduce the amount
of data by block averaging. Using {\tt Nav = (2,2)} effectively quadruples the
area covered by one pixel and therefore reduces number of returned
$q$-positions to 200$\times$200 per coordinate.

Furthermore the use of the algorithms to directly convert angles to reciprocal
space indices $h$, $k$ and $l$ is shown. A function named {\tt Ang2HKL}, which
is called together with the orientation matrix to directly convert the
experimental angles to the indices $h$, $k$, $l$ instead of the momentum
transfer in units of~\AA$^{-1}$. Instead of specifying the orientation matrix
one can also use the crystallographic orientation of the sample to determine
the matrix $\mat U$, which is the default if no $\mat U$-matrix is explicitly
given. Special algorithms also exist for determining the matrix $\mat B$ from
unit cell vectors. In the example the matrix $\mat B$ is determined directly
from the built in Material properties of silicon.  In the present case
(horizontal diffraction) the momentum transfer for the area detector has mainly
a component along negative $y$-direction, whereas the $h$, $k$, $l$ components
have there largest component along $l$.

Experimental angles do of course not need to be entered manually.  They can be
read from several different data formats as mentioned above.  Details are given
in the documentation found online.

\section{rotation matrix for arbitrary axis direction}

Several rotations need to be performed around none primitive directions, e.\,g.
in the case of $\kappa$-goniometers or also when calculating detector tilts. In
such case the rotation axis as well as the angle of rotation are known and
rotation matrices are generated using an formula given in~\cite{Lang1998}. The
components or the rotation matrix $(\mat R)_{ij} = r_{ij}$ for rotation of angle
$\alpha$ around the axis $\vec e$ are given by
\begin{equation}
r_{ij} = e_i e_j + (\delta_{ij} - e_i e_j) \cos\alpha - \epsilon_{ijk} e_k \sin\alpha
\label{eq:rotarb}
\end{equation}
with the Kronecker delta $\delta_{ij}$ and the Levi-Civita symbol $\epsilon_{ijk}$.

     %-------------------------------------------------------------------------
     % The back matter of the paper - acknowledgements and references
     %-------------------------------------------------------------------------

     % Acknowledgements come after the appendices

\ack{Acknowledgements}

We acknowledge the help of our local contact for the measurements performed at
the synchrotron beamline ID01 ESRF Grenoble (T. Sch\"ulli) and the financial
support of the Austrian Science fund (IRON SFB25). DK acknowledges the support
by Austrian Academy of Sciences (DOC-fellowship). Special thanks also to the
early users for their feedback (T.~Etzelstorfer, M.~Keplinger, R.~Grifone, S.~Cecchi).

     % References are at the end of the document, between \begin{references}
     % and \end{references} tags. Each reference is in a \reference entry.

\bibliographystyle{iucr}
\bibliography{iucr}

%\begin{references}
%\reference{Author, A. \& Author, B. (1984). \emph{Journal} \textbf{Vol},
%first page--last page.}
%\end{references}

     %-------------------------------------------------------------------------
     % TABLES AND FIGURES SHOULD BE INSERTED AFTER THE MAIN BODY OF THE TEXT
     %-------------------------------------------------------------------------

     % Simple tables should use the tabular environment according to this
     % model

%\begin{table}
%\caption{Caption to table}
%\begin{tabular}{llcr}      % Alignment for each cell: l=left, c=center, r=right
% HEADING    & FOR        & EACH       & COLUMN     \\
%\hline
% entry      & entry      & entry      & entry      \\
% entry      & entry      & entry      & entry      \\
% entry      & entry      & entry      & entry      \\
%\end{tabular}
%\end{table}

     % Postscript figures can be included with multiple figure blocks

\end{document}